\documentclass{article}
\usepackage{graphicx}

\begin{document}

\title{Wave trains, self-oscillations and synchronization in discrete media}
\author{A. Carpio\thanks{ \tt ana$_-$carpio@mat.ucm.es}\\
Departamento de Matem\'{a}tica Aplicada\\ Universidad Complutense
de Madrid\\ 28040 Madrid, Spain}

\date{ \today   }


\maketitle

\begin{abstract}
We study wave propagation in networks of coupled cells which can
behave as excitable or self-oscillatory media. For excitable media,
an asymptotic construction of wave trains is presented. This
construction predicts their shape and speed, as well as the critical
coupling and the critical separation of time scales for propagation
failure. It describes stable wave train generation by repeated
firing at a boundary. In self-oscillatory media, wave trains persist
but synchronization phenomena arise. An equation describing the
evolution of the oscillator phases is derived.
\end{abstract}

\section{Introduction}
\label{sec:intro}

Understanding wave propagation and self-oscillations in discrete
media is important to predict the behavior of many biological and
physical systems composed of smaller interacting units, such as
cells, nodes, atoms, wells...  Examples abound: propagation of
impulses through myelinated axons \cite{scott2} or cardiac tissue
\cite{beeler}, voltage oscillations in muscle fibers \cite{morris},
charge transport in semiconductor superlattices \cite{bonilla},
crystal growth and interface motion in crystalline materials
\cite{cahn}, etc.

In this paper, we consider an infinite chain of diffusively
coupled cells described by a fast variable $v_n$ and a slow variable
$w_n$:
\begin{eqnarray}
{dv_n \over dt}= D (v_{n+1}-2v_n+v_{n-1}) + f(v_n,w_n), \label{int1}\\
{dw_n \over dt}= \lambda g(v_n,w_n). \label{int2}
\end{eqnarray}
Typically, $v_n(t)$ represents the ratio of the membrane potential
to some reference equilibrium potential and $w_n(t)$ the fraction
of open channels for some type of ions at the $n$-th site.
The parameter $D>0$ measures the strength of the coupling and
$\lambda >0$ represents the ratio  of the time scales for $v_n$
and $w_n$. $\lambda$ must be small enough for the two variables to
evolve in different time scales.
We assume that the nullclines $g(v,w)=0=f(v,w)$ intersect at one point,
the equilibrium of the system. For $w$ fixed, the source $f(v,w)$ is
bistable. System (\ref{int1})-(\ref{int2}) displays excitable or
self-oscillatory behavior depending on whether the equilibrium state
is stable or unstable.

In the excitable case, periodic disturbances in the medium lead to
propagation of signals in the form of wave trains.
To obtain analytic information on those waves, it is quite tempting
to replace the discrete system (\ref{int1})-(\ref{int2}) by its continuum
limit $D\rightarrow \infty$:
\begin{eqnarray}
{\partial v  \over \partial t}= {\partial^2 v \over \partial x^2} + f(v,w),\quad
{\partial w  \over \partial t}= \lambda g(v,w). \label{cont}
\end{eqnarray}
There is a good mathematical reason not to do that: propagation failure.
The discrete system has a threshold coupling for wave propagation whereas
the continuous model allows wave propagation at all coupling strengths.
Constructing wave train solutions of discrete systems is not easy.
An asymptotic understanding of these solutions and their numerical
construction have to go hand in hand. In an infinite chain, most
initial conditions will either fail to propagate or evolve into a
solitary pulse. Only very particular initial conditions evolve
into a wave train, as we will show later in this paper after
explaining the asymptotic construction thereof. Simpler pulse
solutions were constructed asymptotically in \cite{cbfhn} for the
discrete FitzHugh-Nagumo model. For continuous
models, asymptotic descriptions of pulse waves are given in
\cite{keener3}. Unlike continuous model equations, travelling waves
in discrete media do not solve a system of ordinary differential
equations. Instead, a more complex system of differential-difference equations
needs to be analyzed. Rigorous results are only available for particular
sources and traveling pulses. Tonnelier \cite{tonnelier} found
explicit expressions of  pulses for piecewise linear sources.
Existence of discrete pulses has been proved for a restricted class
of nonlinear sources \cite{hastings}. Conditions for propagation
failure were discussed in \cite{sleeman,erneux,cbfhn}.

In this paper, we give an asymptotic construction of wave
trains in discrete excitable systems. Solitary pulses are a particular
case of a wave train with infinite spatial period and maximum speed.
We study the dynamic equations (\ref{int1})-(\ref{int2})  for general
smooth nonlinear source terms.
Experimental values of $D$ (for instance in nerve models)
are typically small. Thus, we focus in the highly discrete case:
$D \ll 1$.
We find a one-parameter family of stable wave trains and we predict
their speed, period and shape, as well as the critical coupling
for propagation failure. Near the critical coupling,
the profiles develop steps as shown by Figure \ref{figura1}(b) and
Figure \ref{figura5}. Propagation is `saltatory': it proceeds
by a sequence of successive jumps. The mechanism for propagation
failure as the coupling decreases is reminiscent of the depinning
transitions for wave fronts described in \cite{cb}.
Propagation also fails when the parameter $\lambda$ characterizing
the ratio of time scales is no longer small. Our asymptotic construction
yields sharp upper an lower bounds on  the critical value of $\lambda$
below which propagation fails, $\lambda_{c}(D)$, as a function of the
coupling $D$. For $D$ small, we also obtain an asymptotic prediction which
describes accurately numerical measurements. Technically, we have greatly
improved the rough bounds contained in \cite{hastings}, derived under
very restrictive assumptions on the nonlinearities.
Bounds on the critical separation of scales are unknown in the continuum
case.

When the medium is self-oscillatory, the mechanism leading to
periodic wave train generation is completely different. Spatially
uniform profiles corresponding to stable limit cycles of the system:
\begin{eqnarray} {dv  \over dt}=
f(v,w), \; {dw \over dt} = {\lambda}  g(v,w),
\label{ode} \end{eqnarray}
play a key role and introduce new features such as synchronization.
For small couplings, we derive an equation for the time evolution of
oscillator phases.

The paper is organized as follows. In section \ref{sec:trains} we
give an asymptotic description of periodic wave trains for excitable systems.
Section \ref{sec:numerics} discusses their numerical stability and
selection via initial and boundary conditions. We use the
FitzHugh-Nagumo (FHN) and the more complicated Morris-Lecar
(ML) dynamics as examples in our numerical calculations.
Section \ref{sec:propagation} analyzes the impact of discreteness
on propagation failure. We predict the critical coupling and the critical
separation of time scales for propagation failure as a function of the
coupling. Section \ref{sec:oscillatory} is devoted to self-oscillatory
behavior and synchronization. Section \ref{sec:conclusions} contains our
conclusions.

\section{Wave trains and pulses in excitable media}
\label{sec:trains}

\begin{figure}
\begin{center}
\includegraphics[width=8cm]{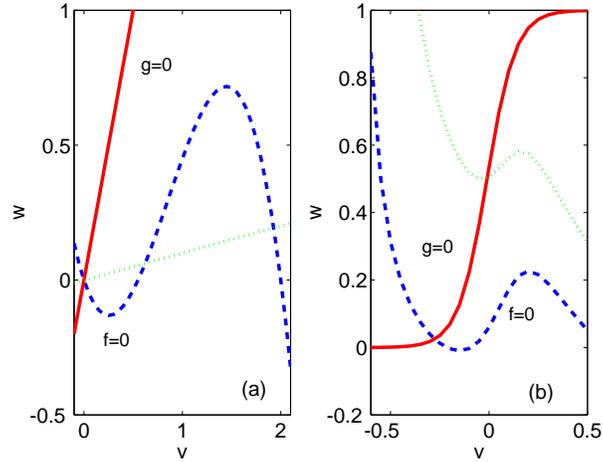}
\caption{ (a) FHN with $a=0.3$ and $b=0.5$ (solid), $b=10$ (dotted);
(b)  ML with $I=0.0625$ (dashed), $0.375$ (dotted).}
\label{figura2}
\end{center}
\end{figure}

Figure \ref{figura2} depicts the nullclines for system
(\ref{int1})-(\ref{int2}) and the sources we have used
in the numerical tests.  We assume that the source $f(v,w)$ is
a cubic function of $v$ with three zeros $z_1(w)<z_2(w)<z_3(w)$, the
first and the third of which are stable for the reduced equation
(\ref{int1}) with $w_n=w$. The nullclines must intersect in the
first or third branch of the cubic to guarantee that
(\ref{int1})-(\ref{int2}) describe an excitable system.
Intersections of the nullclines in the middle branch  of the
cubic correspond to unstable stationary states and give rise
to self-oscillatory dynamics.
To fix ideas, we assume in this section that the nullclines
intersect at a point $(v^0,w^0)$ located in the first branch of
the cubic.

Experimental values of $D$ (for instance in nerve models) are
typically small. Thus, we focus in highly discrete problems with
$D \ll 1$. As shown in Figure  \ref{figura12}, a wave train
resembles a sequence of pulses.
After a short transient the system relaxes to
a traveling wave: $v_n(t)=V(n-ct)$, $w_n(t)=W(n-ct)$, in which $V$
and $W$ are periodic functions with the same period $L$.
This defines the wave profiles $V(z)=v_n({n-z\over c})$ and
$W(z)=w_n({n-z\over c})$, $z=n-ct$. Neighboring
points describe the same trajectory with a time delay ${1\over c}$:
$v_{n+1}(t)=v_n(t-{1\over c}),$ $w_{n+1}(t)=w_n(t-{1\over c})$.

The profiles and speeds solve an eigenvalue
problem for a system of differential-difference equations:
\begin{eqnarray}\begin{array}{l}
-c V_z(z)\!=\! D\!(V(z\!+\!1)\!-\! 2V(z)\!+\!V(z\!-\!1))
\! +\! f(V(z),W(z)), \\
-c W_z(z)\!=\! \lambda g(V(z),W(z)),
\end{array}\label{p0}  \end{eqnarray}
with $V(z+L)=V(z)$, $W(z+L)=W(z)$.
Equations (\ref{p0}) are hard to solve except in very particular
situations, such as  piecewise linear $f$ and linear $g$ \cite{tonnelier}.
Is there an alternative way to extract practical information
on the waves? We show below  how to use asymptotic methods
to predict the ranges of the parameters for which we expect
stable wave trains to be formed, the shape of their profiles,
their spatial period $L$ and speed $c$.

\subsection{The reduced bistable equation}
\label{sec:bistable}

As Figure \ref{figura12} shows,  a wave train for $v_n$ is a sequence
of sharp interfaces separated by smooth transitions.
At the interfaces, the recovery $w_n$ is almost constant. Thus, the first
step in our asymptotic construction is the study of wave front
propagation in the bistable equation:
\begin{eqnarray}
{dv_n \over dt}= D (v_{n+1}-2v_n+v_{n-1}) +f(v_n,w). \label{p1}
\end{eqnarray}
Choosing $w$ in a neighborhood $(w_-,w_+)$ of the equilibrium value
$w^0$, the source
$f(v,w)$ has three zeros $z_1(w)<z_2(w)<z_3(w)$, the first and the
third of which are stable. The parameter $w$ controls the degree
of asymmetry of the source, as shown in Figure \ref{figura3}(b).
In particular, it changes the ratio of the area $A_{12}$ above $f$
between $z_1$ and $z_2$ and the area $A_{23}$ below $f$ between
$z_2$ and $z_3$. When the cubic source $f(v,w)$ is asymmetric
enough, depending on $D$, wave fronts $v_n=V(n-ct)$ joining the
two stable zeros $z_1(w)$ and $z_3(w)$ propagate
\cite{ejam,cb,fath,mallet,zinner}. Generically, there are two
threshold values $w_{c-}(D)$,$w_{c+}(D)$ such that:
\begin{itemize}
\item If $w\in (w_-,w_{c-}(D))$, wave fronts decreasing from $z_3(w)$
to $z_1(w)$ (resp. increasing from $z_1(w)$ to $z_3(w)$) move to
the right (resp. left) with speed $c=c_+(w,D)$.
\item If $w\in (w_{c+}(D),w_+)$, wave fronts increasing from $z_1(w)$
to $z_3(w)$ (resp. decreasing from $z_3(w)$ to $z_1(w)$) move to
the right (resp. left) with speed $c=c_-(w,D)$.
\end{itemize}
In the first case, $A_{23}>A_{12}$. In the second, $A_{23}<
A_{12}$. If $w \in [w_{c-}(D),w_{c+}(D)]$, fronts are pinned and $c=0$.
The size of this pinning interval grows as $D$ decreases.

Let us recall some facts about the structure of wave front solutions,
which will be needed in Section \ref{sec:propagation}.
For small $D$, the stable stationary front  $s_n$
with  $w=w_{c-}(D)$  has the following structure: two
essentially constant tails $s_n\sim z_3(w), n<0,$ and $s_n \sim
z_1(w), n>0,$ joined by an {\it active} point  $s_0$, which takes values
in the interval $ (z_1(w), z_2(w))$. When $w$  is near the threshold
$w_{c-}(D)$, the travelling wave front profiles are staircase-like.
 While the wave fronts move at a constant speed $c$, the motion
of points $v_n(t)$ belonging to these wave fronts is saltatory:
their spatial profiles $v_n$ stay near the shifted static
configurations $s_{n-l}$, $l\in  Z\!\!\!\! Z$,  for a time
${1\over c}$, until the $l$-th point $v_l(t)$ (the {\it active point})
jumps from $s_0$ to $z_3(w)$ and $v_{l+1}(t)$ (the {\it next active point})
jumps from  $z_1(w)$ to $s_0$, making the front advance one position.
Increasing $D$, we find a similar situation with an increasing number of
{\it active} points
filling the gap between the constant tails in the spatial profiles and an
increasing number of steps in the temporal profiles. As $w$ grows, the
profiles and their motion become smoother, see \cite{cb,fath}.

The speed of the wave fronts can be predicted in two limits. In
the depinning transition $w\rightarrow w_{c\pm}(D)$, we find
$c\approx {\rm sign}(w-w_{c\pm}(D)) \sqrt{\alpha\beta(w-w_{c\pm}(D))}/\pi$,
see \cite{cb} for details. For highly discrete problems with $D\ll 1,$
fronts can  move only when  $w\approx w_{c\pm}(D) $ and the approximation
for the speed provided by the analysis of the depinning transition
always holds. This scaling changes when $f$ is a
piecewise linear function, as shown in \cite{fath}.
In the continuum limit $D\gg 1$, the wave front velocities are a
correction of the wave front velocities $c_0$ for a spatially
continuous problem $c\approx \sqrt{D} c_0 (1-{k c_0^2\over 2D})$,
see \cite{keener2} for details. Out of these two limits,
characterized by either small or large speeds, we compute the
speeds by solving (\ref{p1}) numerically.

The picture of wave front propagation we have described  holds
for any smooth source $f$ when $D$ is small enough.
For larger $D$, we may encounter sources $f$ for which
the pinning interval shrinks to a point \cite{cb}. Finding
a complete characterization of the sources for which  there is no
pinning is an unsolved problem. These anomalous cases are excluded
from our analysis.

\subsection{Asymptotic construction of wave trains}
\label{sec:asymptotics}

\begin{figure}
\begin{center}
\includegraphics[width=8cm]{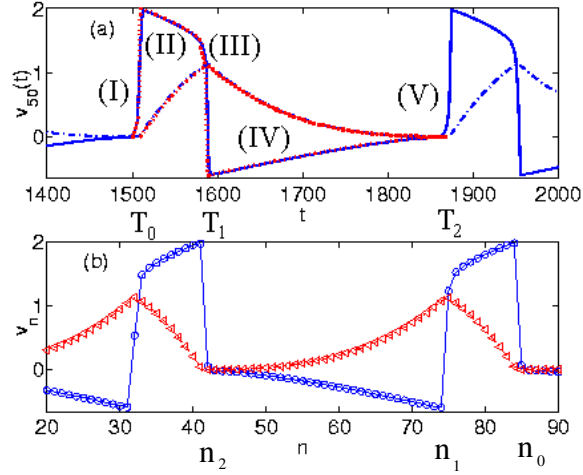}
\caption{Wave train for FHN:
(a) Temporal profiles of $v_n(t)$ (solid) and $w_n(t)$ (dashed-dotted),
compared to their asymptotic reconstruction (dotted);
(b) Spatial profiles of $v_n$ (circles) and $w_n$ (triangles).
Parameter values are  $a=0.1$, $b=0.5$, $D=0.02$,
$\lambda=0.01$, $w=0.002$.}
\label{figura12}
\end{center}
\end{figure}

Having described the reduced dynamics of wave front solutions of bistable
equations in Section \ref{sec:bistable}, we are ready to predict asymptotically
the speed and shape of wave trains for (\ref{int1})-(\ref{int2}) in the limit
 $\lambda \rightarrow 0^+$.

Discrete wave trains have the form $v_{n}(t)=V(n-ct)$, $w_{n}(t)=W(n-ct)$.
 $V$ and $W$ are {\em real functions of a real variable} $z=n-ct$. Note that $z$
 takes on continuous values and not discrete values like the variable $n$.
 Thus $V$ and $W$ are perfectly determined once their speed $c$, period $L$
 and continuous profiles $V(z)$ and $W(z)$, $z \in [0, L]$, are known.
If we fix $n$, the trajectories $v_{n}(t)=V(n-ct)$, $w_{n}(t)=W(n-ct)$ are periodic
in the  continuous variable $t$ and have a well defined temporal period $L/c$.
Thus,  once we compute the speed $c$ and the temporal profiles $v_{n}(t),w_{n}(t)$
as functions of $t$ for fixed $n$,  the wave train profiles $V$ and $W$ and their
period $L$ are known.
 If we fix $t$ and vary the discrete spatial variable $n$, the situation is different.
 The spatial structure of the wave train at each fixed time $t$ is given by $v_{n}=
 V(n-ct)$ and $w_{n}=W(n-ct)$. The points contained in a period satisfy $0 \leq n-ct
 \leq L$, that is, $ct \leq n \leq L+ct$. As time grows, the discrete points 'travel'
 along the continuous wave profile and are transferred from one period to the next.
 The number of integers $n$ one can fit in an interval of length $L$ is the integer part
 of  $L$.

There are no rigorous existence results for wave train solutions of
(\ref{int1})-(\ref{int2}).
Assuming that wave trains do exist, we shall use the
separation of time scales $\lambda$ to find an approximate
reconstruction of the temporal wave train profiles by matched asymptotic
expansions  at zeroth order as $\lambda \rightarrow 0^+$.
A description of matched asymptotic expansions can be found in [18] or in
the more elementary reference [3].

For $D$ small, numerical solutions show that wave train trajectories
should be composed of regions of smooth variation of $(v_n,w_n)$ on
the slow time scale $T=\lambda t,$  separated by sharp interfaces
in which $v_n$ varies rapidly in the fast time scale $t$, see Figure
\ref{figura12}. The reference time in our construction of trajectories will
be the slow time $T.$

For $n$ fixed, we may distinguish four stages in one time period
of the wave train solutions $(v_n(T;\lambda),w_n(T;\lambda))$.
At each stage, a reduced description holds:
\begin{itemize}
\item {\it Leading edge} {\cal (I)} in Figure \ref{figura12} (a).
The variable $v_n$ evolves abruptly in the fast time scale
$\mbox{\~t}={T-T_0\over \lambda} \in(-\infty,\infty)$. To leading order:
\begin{eqnarray}\begin{array}{l}
{dv_n \over d\mbox{\~t}}= D (v_{n+1}-2v_n+v_{n-1}) + f(v_n,w_n), \quad
{dw_n \over d\mbox{\~t}}= 0.
\end{array}\label{p2} \end{eqnarray}
Thus, $w_n=w$ is a constant and $v_n$ is the temporal
profile of the wave front solution of (\ref{p1}) which joins
 $z_1(w)$ and $z_3(w)$. This front propagates with a
speed $c_+(w,D)$ when $w \in (w_-,w_{c-}(D))$.

\item {\it Peak.} This is region {\cal (II)} in  Figure
\ref{figura12} (a).  Here $v_n$ varies smoothly on the slow
time scale $T$. Thus, we may set  $D=0$, obtaining the
reduced problem $f(v_n,w_n)=0,$ ${dw_n \over dT} =  g(v_n,w_n)$.
Matching the peak to the leading edge at zeroth order,
we see that $v_n$ must lie in the third branch
of roots of the cubic function: $v_n=z_3(w_n).$ Then,
for $T_0 < T < T_1(w,\overline{w})$:
\begin{eqnarray}
{dw_n \over dT} = g(z_3(w_n),w_n), \label{p3}\\
T_1(w,\overline{w})=T_0+ \int_w^{\overline{w}}\!{ds \over
g(z_3(s),s)}, \label{l1}
\end{eqnarray}
in which $w_n(T_0)=w$ and $w_n(T_1)=\overline{w}$ (to be calculated
later).

The matching conditions at  $T=T_0$ for the reduced descriptions of
the peak and the leading front are
\footnote{$f(\epsilon)\ll g(\epsilon)$ as
$\epsilon\to 0$ is equivalent to $f=o(g)$ and it means lim$_{\epsilon\to 0}f/g=0$,
see \cite{bender}.}:
\begin{eqnarray}\begin{array}{l}
z_3(w_n^{(II)}(T))-v_n^{(I)}
({T-T_0\over \lambda}) \ll 1, \quad
w_n^{(II)}(T) -w \ll 1,
\end{array}\nonumber\end{eqnarray}
as $\lambda\to 0 +$ in the overlap region $\lambda \ll T-T_0 \ll 1$.
The superscripts (I), (II) ... refer to the region and the reduced
description we are using.
The uniform approximations of the profiles
in stages (I)-(II) are:
\begin{eqnarray}\begin{array}{l}
v_n^{unif} =
v_n^{(I)}({T-T_0\over \lambda}) + z_3(w_n^{(II)}(T)) - z_3 (w), \quad
w_n^{unif}=w_n^{(II)}(T).
\end{array}\nonumber\end{eqnarray}

\item {\it Trailing wave front.} The peak is followed by another
region of fast variation of $v_n$, corresponding to
{\cal (III)} in  Figure \ref{figura12} (a).
Again, (\ref{p2}) holds to leading order,  on the new time scale
$\mbox{\~t}={T-T_1\over \lambda}\in(-\infty,\infty)$. Thus,
$w_n=\overline{w}$ is a constant and $v_n$ is the temporal profile of
the wave front solution of (\ref{p2}) with $w_n=\overline{w}$ which
joins $z_3(\overline{w})$ and $z_1(\overline{w})$. This front
propagates with a speed $c_-(\overline{w},D)$ when $\overline{w}
\in (w_{c+}(D),w_+)$. If the wave train is to move rigidly,
$\overline{w} $ must be selected in such a way that the trailing front
travels with speed $c_-(\overline{w},D)=c_+(w,D)=c$.

The matching conditions for the reduced descriptions of the peak
and the trailing front at $T=T_1$ are:
\begin{eqnarray}\begin{array}{l}
z_3(w_n^{(II)}(T))-v_n^{(III)}({T-T_1\over \lambda}) \ll 1, \quad
-w_n^{(II)}(T) + \overline{w}  \ll 1,
\end{array}\nonumber\end{eqnarray}
as $\lambda\to 0 +$ in the overlap region $\lambda \ll T_1-T \ll 1$.
The uniform approximations of the profiles in regions (I)-(III) are:
 \begin{eqnarray}\begin{array}{l}
v_n^{unif} = v_n^{(III)}({T-T_1\over \lambda}) +
v_n^{(I)}({T-T_0\over \lambda})
+ z_3(w_n^{(II)}(T)) - z_3 (w) - z_3 (\overline{w}), \\
 w_n^{unif}=w_n^{(II)}(T).
\end{array}\nonumber\end{eqnarray}

\item {\it Tail.} The last part of the period is region {\cal (IV)}
in  Figure \ref{figura12} (a). Here, $v_n$ varies smoothly
on the slow time scale $T$, and $D$
can be neglected. The reduced problem is $f(v_n,w_n)=0,$
${dw_n \over dT} =  g(v_n,w_n)$.
Matching the tail to the trailing wave front at zeroth order,
we see that $v_n$ must lie in the first branch of roots:
$v_n=z_1(w_n).$ Then, for $T_1 < T < T_2(w,\overline{w})$:
\begin{eqnarray}
{dw_n \over dT} = g(z_1(w_n),w_n), \label{p4} \\
T_2(w,\overline{w})= T_1+ \int_{\overline{w}}^w \!{ds \over  g(z_1(s),s)},
\label{l2}
\end{eqnarray}
in which $w_n(T_1)=\overline{w}$ and $w_n(T_2)=w=w_n(T_0)$, if the wave
train is to move rigidly.
In this way, we  conclude the description of one temporal period
and guarantee that the end of its tail can be matched to the following
(identical) period.
Notice that the integral (\ref{l2}) diverges when $w\geq w^0 $
due to a singularity at the equilibrium value $w^0$. Thus, the range of
$w$ for which we can find wave trains is restricted to
$(w^0,w_{c-}(D)).$

The matching conditions for the reduced descriptions of the tail
and the trailing front at $T=T_1$ are:
\begin{eqnarray}\begin{array}{l}
v_n^{(III)}({T-T_1\over \lambda}) - z_1(w_n^{(IV)}(T)) \ll 1, \quad
-w_n^{(IV)}(T) + \overline{w}  \ll 1,
\end{array}\nonumber\end{eqnarray}
as $\lambda\to 0 +$ in the overlap region $\lambda \ll T- T_1 \ll 1$.
The uniform approximations of the profiles in regions (III)-(IV) are:
 \begin{eqnarray}\begin{array}{l}
v_n^{unif} = v_n^{(III)}({T-T_1\over \lambda}) + z_1(w_n^{(IV)}(T)) -
z_1 (\overline{w}), \quad w_n^{unif}=w_n^{(IV)}(T).
\end{array}\nonumber\end{eqnarray}
In a similar way, the end of the tail should be matched to the leading
edge (region (V)) of the next period at $T_2$.

\end{itemize}

A uniform approximation to the temporal profile in a whole period
is obtained as follows. Let $\chi(a,b)$ be the indicator function, equal
to $1$ if $a \leq T \leq b$ and $0$ otherwise.
Then, for $T_0\leq T \leq  T_2$:
\begin{eqnarray}\begin{array}{l}
v_n^{unif}(T) = \big[
v_n^{(III)}({T-T_1\over \lambda}) \!+\! v_n^{(I)}({T-T_0\over \lambda})
\!+\! z_3(w_n^{(II)}(T)) \!-\! z_3 (w) \!-\! z_3 (\overline{w}) \big]
\chi(T_0,T_1)  \\
\quad \quad \quad \;\;
+\big[v_n^{(III)}({T-T_1\over \lambda}) \!+\! v_n^{(V)}({T-T_2\over \lambda})
\!+\! z_1(w_n^{(IV)}(T)) \!-\! z_1 (w) \!-\! z_1 (\overline{w}) \big]
\chi(T_1,T_2),\\
w_n^{unif}(T)=w_n^{(II)}(T) \chi(T_0,T_1) + w_n^{(IV)}(T) \chi(T_1,T_2).
\end{array}\nonumber\end{eqnarray}

This construction yields a family of wave trains parametrized by
the values $w\in (w^0,w_{c-}(D))$   for which there exists
$\overline{w}=\overline{w}(w) \in (w_{c+}(D),w_+)$ satisfying
$c_-(\overline{w},D)=c_+(w,D)=c$.

In Figure \ref{figura12} (a), we have superimposed the approximate
temporal profile provided by our matched asymptotic expansions and
the numerically calculated  profile. They are indistinguishable.
Notice that, for small $D$, the leading and
trailing wave fronts can be accurately computed
using one active point (see section 2.1 or \cite{cb}):
\begin{eqnarray}
{dv_n \over d\mbox{\~t}}= D (z_3(w)-2v_n+z_1(w)) + f(v_n,w), \label{p2bis} \\
{dv_n \over d\mbox{\~t}}= D (z_3(\overline{w})-2v_n+z_1(\overline{w})) +
f(v_n,\overline{w}).
 \end{eqnarray}
In the true time scale $t={T\over \lambda}$, the temporal period
of the wave train is asymptotic to
$\big( T_2(w) -T_0\big)/ \lambda$
(the duration of the fast stages (I) and (III) is ignored).

Having calculated the temporal profiles $v_{n}(t),w_{n}(t)$ of a  wave train
and its velocity $c$, the wave train profiles $V$, $W$ are known:
$V(z)=v_n({n-z\over c})$ and $W(z)=w_n({n-z\over c})$, $z=n-ct$.
Their period is now asymptotic to $c_+(w) \big( T_2(w) -T_0 \big)/ \lambda$.
The number of points $L(w)$ in the wave length of the wave train is essentially
the integer part of the wave length. Figure \ref{figura12} (b)
illustrates that, at a fixed time $t$:
\begin{itemize}
\item the leading front of one spatial period is located at a
point $n_0$, where $w_{n_0}(t)\sim w$,
\item its trailing front is located at a point $n_1$, such that
$w_{n_1}(t)\sim \overline{w}$,
\item the end of our spatial period and the beginning of the next
one is located at a point $n_2$, where $w_{n_2}(t)\sim w$.
\end{itemize}
Then, $L(w)=n_0-n_2$ is approximately the integer part of
$c_+(w) \big( T_2(w) -T_0  \big)/ \lambda$. Notice that the number of
points inside the leading and trailing fronts of each period can
be ignored in the discrete limit $D\ll 1$. The number  of points
in the peak $L_1(w)=n_0-n_1$   is asymptotic to the integer part
of $c_+(w) \big(T_1(w)-T_0\big)/ \lambda$ and the number of points
in the tail $L_2(w)=n_1-n_2$ is asymptotic to the integer part
of $c_+(w) \big(T_2(w)-T_1(w)\big)/ \lambda$.
Our construction is consistent only
when $L_1(w)\geq 1$. This  provides an estimate of the critical
value $\lambda_c(D)$ above which propagation fails.
Nevertheless, near the thresholds for propagation failure, the
approximation described in this section has to be corrected.
This is explained in more detail in Section \ref{sec:propagation}.

\subsection{Solitary pulses}
\label{sec:pulses}

A wave train becomes a traveling pulse when its spatial period tends
to infinity, which occurs as $w\rightarrow w^0$. Pulses are the
fastest waves of the family.
In the asymptotic description of the temporal profile of a pulse we
distinguish five regions:
\begin{itemize}
\item  In front of the pulse, the profile is at equilibrium: $v_n\sim
v^0$, $w_n\sim w^0$.
\item The leading edge of the pulse is a wave front solution
of  (\ref{p2}) with $w_n=w^0$, which joins
$z_1(w^0)=v^0$ and $z_3(w^0)$. This front propagates with a
definite speed $c=c_+(w^0,D)$.
\item In the transition between fronts, $v_n=z_3(w_n)$,
${dw_n \over dT} =  g(z_3(w_n),w_n)$ and $w_n$ evolves from
$w^0$ at $T_0$ to $\overline{w}^0$ at $T_1$.
\item The trailing wave front is the wave front solution of
(\ref{p2}) with $w_n=\overline{w}^0$, which joins
$z_3(\overline{w}^0)$ and $z_1(\overline{w}^0)$.
$\overline{w}^0$ is now selected in such a way
that this front travels with speed
$c=c_-(\overline{w}^0,D)=c_+(w^0,D)$.
\item In the pulse tail,  $v_n=z_1(w_n)$,
${dw_n \over dT} =  g(z_1(w_n),w_n) $ and $w_n$ evolves
from  $\overline{w}^0$ to $w^0$.
\end{itemize}

The number of points between fronts $L_1(w^0)$ is again asymptotic
to the integer part of of $c_+(w^0) (T_1(w^0)-T_0)/ \lambda$.
The number of points in the tail is infinite, since the integral
$ \int_{\overline{w}^0}^{w^0}  {ds \over g(z_1(s),s)}$ diverges.
However, we can   predict how long does it take for the
tail to get sufficiently close to $v^0=z_1(w^0)$ by using
$\int_{\overline{w}^0}^{w^0-\epsilon} {ds \over
g(z_1(s),s)},$ with $ \epsilon>0$ instead.

\section{Periodic firing of waves at boundaries}
\label{sec:numerics}

When an excitable medium is periodically excited, we expect
propagation of signals in form of wave trains.  We will
see  here that the periodic wave train solutions constructed in
Section \ref{sec:trains} allow to describe propagation due to
periodic firing at a boundary in the spatially discrete
FHN and ML models.

\subsection{FitzHugh-Nagumo}
\label{sec:fhn}

The FitzHugh-Nagumo (FHN) model is a crude simplification of
(\ref{int1})-(\ref{int2}) that sets $f(v,w)=h(v)-w$ with $g$
linear. Typically, $h$ is chosen to be a cubic polynomial,
say $h(v)=v(v-a)(2-v)$. We get:
\begin{eqnarray}
{dv_n \over dt}\!=\! D (v_{n+1}\!-\!2v_n\!+\!v_{n-1}) \!+\!
 v_n(v_n\!-\!a)(2\!-\!v_n)\!-\!w_n, \label{fh1} \\
{dw_n \over dt}= \lambda (v_n- bw_n). \label{fh2}
\end{eqnarray}

\begin{figure}
\begin{center}
\includegraphics[width=8cm]{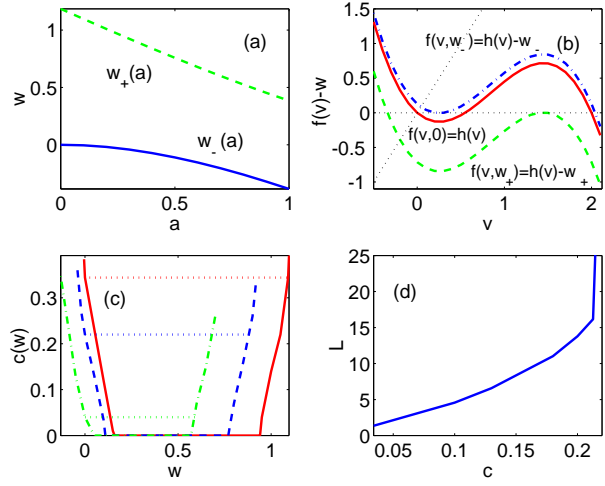}
\caption{For FHN: (a) Critical curves for cubic shape; (b) Change of
symmetry in the source as $w$ increases from $w_-$ to $w_+$;
(c) Speed curves  when $D=0.1$ for $a=0.55$ (solid), $a=0.3$
(dashed), $a=0.1$ (dashed-dotted); (d) Dispersion
law $L(c)$ for $D=0.1$ and $\lambda=0.05$.}
\label{figura3}
\end{center}
\end{figure}

When $a\in (0,1)$, this system is excitable provided $b$ is not
too large, as illustrated in Figure \ref{figura2}(a). The unique
stationary state is $(v^0,w^0)=(0,0)$. As Figure \ref{figura3}(b)
shows, the source $v(v-a)(2-v)-w$ is bistable when $w \in(w_-(a),
w_+(a))$. The curves $w_-(a)$, $w_+(a)$ are plotted in Figure
\ref{figura3}(a). The reduced equation (\ref{fh1}) with $w_n=w$ has
decreasing (resp. increasing) traveling wave front solutions moving with
speed  $c_+(w)$ (resp. $c_-(w)$) when $w\in (w_-(a),w_{c-}(a,D))$
(resp. $w\in (w_{c+}(a,D),w_+(a))$). Figure \ref{figura3}(c) depicts
the speed curves $c(w)$ for different values of $a$. The left
branches represent $c_+(w)$ and the right branches $c_-(w)$. The
horizontal dotted lines join $c_+(0)$ with $c_-(\overline{w}^0)$,
selecting the value $\overline{w}^0$ for the trailing front in the
construction of pulses.

\begin{figure}
\begin{center}
\includegraphics[width=8cm]{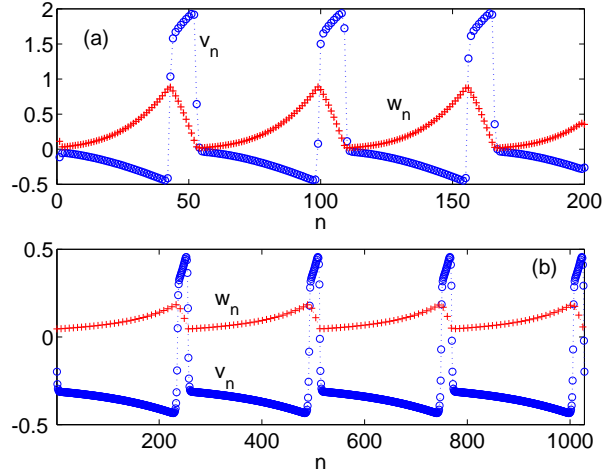}
\caption{(a) Wave train for FHN when  $D=0.1$, $a=0.3$
$\lambda=10^{-2}$, $w=0.02$, $c=0.18$; (b) Wave train for
ML with $D=1$, $\lambda=10^{-3}$,  $w=0.045$, $c=0.54$.
} \label{figura4}
\end{center}
\end{figure}

Wave trains are found when $w\in (0,w_{c-}(a,D))$. They are slower
than pulses. Figure \ref{figura4}(a) shows a wave train for
$a=0.3$, $w=0.02$ and $\lambda=0.01$. It has been generated by solving
(\ref{fh1})-(\ref{fh2}) with initial data $v_n(0)=2$, $n=1,...,10$,
$v_n(0)=0$, $n>10$, $w_n(0)=0.01$ and a periodic excitation at the
left boundary. More precisely, we have used boundary conditions of the
form:
\begin{equation}
v_0(t)= \left\{\begin{array}{ll}
2 & t\in [0,   \rho T]+kT, \\
0 & t\in (\rho T,    T]+kT,
\end{array}\right. \label{b1}
\end{equation}
with $k\in I\!\!N$, $0< \rho <1$. Our theory predicts $c=0.18$,
$L_1=9$ and $L_2=45$, close to the numerically measured values $c=0.175$,
$L_1=10$, $L_2=46$. For $w=0.002$, our theory predicts
$c=0.2$. When $\lambda=0.05$, $L_1=2$ and $L_2=14$. Numerically, we
find $c=0.16$, $L_1=1-2$, $L_2=13-14$.

We have plotted the dispersion relation  $L(c)$ ($L(w)$ against
$c_+(w)$) for our asymptotic family of wave trains in Figure
\ref{figura3}(d). Note that $L'(c)>0$. Our numerical tests with
periodic boundary conditions starting from initial conditions
which are near the shape asymptotically predicted for these wave
trains suggest their stability. For continuous systems like
(\ref{cont}), Maginu \cite{maginu} proved stability of wave trains
under the condition $L'(c)>0$. Our numerical experiments suggest
that a similar stability criterion could be established for wave
trains in discrete systems.

\subsection{Morris-Lecar dynamics}
\label{sec:morris}

The non dimensional Morris-Lecar (ML) model \cite{morris} is:
\begin{eqnarray}
{dv_n \over dt}=D(v_{n+1}-2v_n+v_{n-1}) + f(v_n,w_n) - 2I, \label{rml1} \\
{dw_n \over dt}={\lambda} \cosh({v_n-{V}_3 \over 2 {V}_4})
\big[1 + \tanh({v_n-{V}_3\over {V}_4}) - 2w_n\big],  \label{rml2}
\end{eqnarray}
where  the index $n$ denotes the $n$-th site and:
\begin{eqnarray}
f(v,w) =  2w (v- {V}_K ) + 2 {g}_L (v- {V}_L )
+{g}_{Ca} \big[1+\tanh({v-{V}_1 \over {V}_2 })\big]
(v-1). \label{apb2}
\end{eqnarray}
$v_n$ is the ratio of membrane potential to a reference potential and
$w_n$ is the fraction of open $K^+$ channels.  The time scale is
${\overline{g}_K \over 2 C_m}$, $\overline{g}_K$ being the $K^+$
conductance and $C_m$ the membrane capacitance.
System (\ref{rml1})-(\ref{rml2}) is a reduced version of the
full Morris-Lecar model \cite{morris}, which involves one more
fast variable $m_n$.
Typical values for the  parameters can be taken from experiments
\cite{morris}. In our numerical tests we have used:
\begin{eqnarray}
\begin{array}{|c|c|c|c|c|c|c|c|c|}
\hline
 g_{Ca} &  g_L &  I  &  V_K  &  V_L  &   V_1 & V_2  &  V_3 &  V_4\\
\hline
0.5 & 0.25 & 0.0625 & -0.7 & -0.5 & 0.1 & 0.15 & -0.01 & 0.145\\
\hline
\end{array}\nonumber
\end{eqnarray}

\begin{figure}
\begin{center}
\includegraphics[width=8cm]{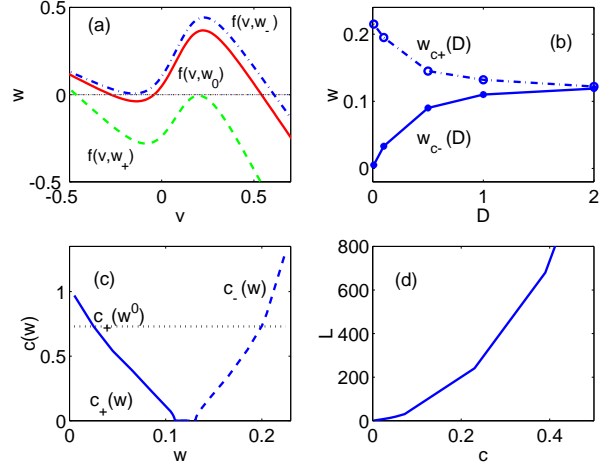}
\caption{For ML:
(a) Change of shape in $f(v,w)$;
(b) Pinning thresholds $w_{c-}(D)$ and $w_{c+}(D)$;
(c) Speed curves $c_-(w)$ and $c_+(w)$  with $D=1$;
(d) Dispersion law $L(c)$ for $D=1$ and $\lambda=10^{-3}$.}
\label{figura6}
\end{center}
\end{figure}

System (\ref{rml1})-(\ref{rml2}) exhibits a rich dynamical behavior
depending on its nullclines. Figure \ref{figura2}(b) shows two possibilities.
For $I=0.375$, the system develops self-oscillations.
This behavior will be discussed in Section \ref{sec:oscillatory}.
For $I=0.0625$, the system is excitable. Figure \ref{figura6}(a) illustrates
the change of symmetry in the source $f(v,w)$ as $w$ increases from
$w_-$ to $w_+$. The thresholds for wave front propagation $w_{c-}(D)$ and
$w_{c+}(D)$ are depicted in Figure \ref{figura6}(b). Note that the size of
the pinning interval grows as $D$ decreases. The curves $c_-(w)$ and
$c_+(w)$ representing the speeds of the wave front solutions of
the reduced bistable equation for this problem with $D$ fixed are
shown in Figure \ref{figura6}(c). We have plotted the dispersion relation
$L(c)$ ($L(w)$ against $c_+(w)$) for our asymptotic family of wave trains
in Figure \ref{figura6}(d). Again, $L'(c)>0$ and this branch of wave trains
is expected to be stable for the dynamics (\ref{rml1})-(\ref{rml2}).

We have generated a variety of wave trains using the asymptotic prediction
as initial datum and periodic boundary conditions. These numerical solutions
agree reasonably well with the asymptotic description given in Sections
\ref{sec:asymptotics}.
Figure \ref{figura4}(b) shows a  wave train for $D=1$, $\lambda=10^{-3}$,
$w=0.045$, $c=0.54$.
For $w=0.045$, the predicted speed is $c=0.54$. When $\lambda=10^{-3}$,
we predict a peak width $L_1 \sim 19$ and a tail width $L_2\sim 256$.
The numerically
measured parameters are $c\sim 0.55$, $L_1 \sim 18$ and $L_2 \sim
250$.

The basin of attraction of wave trains is rather small and their existence
depends crucially on the boundary conditions we employ. There are no wave
trains if we replace the periodic boundary conditions by Dirichlet or Neumann
boundary conditions. With periodic boundary conditions, wave trains fail
to be formed if $w_n(0)$ is near the equilibrium value $w^0$. In such a
case, the initial condition evolves toward a solitary pulse instead of a wave
train.

Pulses are more robust. They can be generated using different
initial data and boundary conditions. For instance, we can chose
step-like initial data $v_n(0)$, climbing from  $v^0$ to a value
close to $z_3(w^0)$. For $w_n(0)$ we may take a perturbation of
$w^0$, as long as $w_n(0)$ is near the value  $\overline{w}^0$ in
the trailing front of the step. These initial conditions evolve into
pulses, regardless of whether we are using Dirichlet or periodic
boundary conditions. When $D=0.1$, our asymptotic construction
predicts pulses with speed $c \sim 0.075$. Setting
$\lambda=10^{-4}$,  the predicted peak width is
$L_1\sim 35$. The tail reaches the value
$w^0-\epsilon$, $\epsilon=10^{-2}$, for $L_{\epsilon}\sim 455$.
Again, these values describe accurately the numerically constructed
pulses.

\begin{figure}
\begin{center}
\includegraphics[width=8cm]{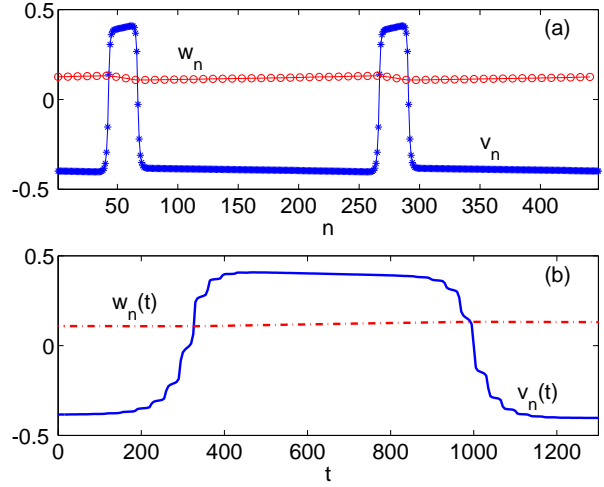}
\caption{Slow wave trains for Morris-Lecar:
(a) spatial profile at a fixed time $t$,
(b) detail of the steps in the temporal profile.
Parameter values are $D=1$, $\lambda=10^{-5}$, $w=0.108$ and
$c=0.04$.}
\label{figura1}
\end{center}
\end{figure}

We observe that the profiles of wave trains and pulses differ slightly
depending on whether their speeds are large or small.
Fast waves have smooth temporal profiles and advance smoothly.
Slow waves develop a sequence of steps in the leading and trailing
fronts. Their motion is 'saltatory'. Figure \ref{figura1} shows the spatial
and temporal profiles of a slow wave train when $D=1$, $\lambda=10^{-5}$,
$w=0.108$ and $c=0.04$. The values $w$ and $\overline{w}$ are close to the
thresholds $w_{c-}(D)$ and $w_{c+}(D)$, that is, near the depinning transitions
\cite{cb} for the bistable equation (\ref{p1}). This is the reason for the
appearance of the steps that can be appreciated in Figure \ref{figura1}(b).
Wave trains and pulses are always 'slow' when $D$ is near the threshold
$D_c$ for propagation failure.
For our parameter values,  $w^0$ is  near the threshold $w_{c-}(D)$ when
$D\sim 0.076$, and the profile of the pulses and wave trains are staircase
like.

\section{Propagation failure}
\label{sec:propagation}

\begin{figure}
\begin{center}
\includegraphics[width=8cm]{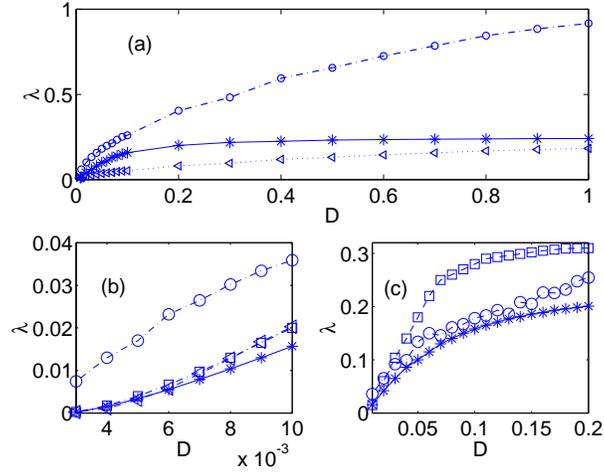}
\caption{ Numerically measured $\lambda_c(D)$ (asterisks) compared to:
(a) the upper bound $\lambda_c(D)^+$ (circles) and the lower bound
$\lambda_c(D)^-$ (triangles);
(b) and (c), the  upper bound $\lambda_c(D)^+$ (circles), the
numerically measured $\lambda_0(D)$ (squares) and its asymptotic
prediction (\ref{hd3}) (triangles). Parameter values for FHN are
$a=0.1$, $b=0.5$, $k=5$.}
\label{figura8}
\end{center}
\end{figure}

\begin{figure}
\begin{center}
\includegraphics[width=8cm]{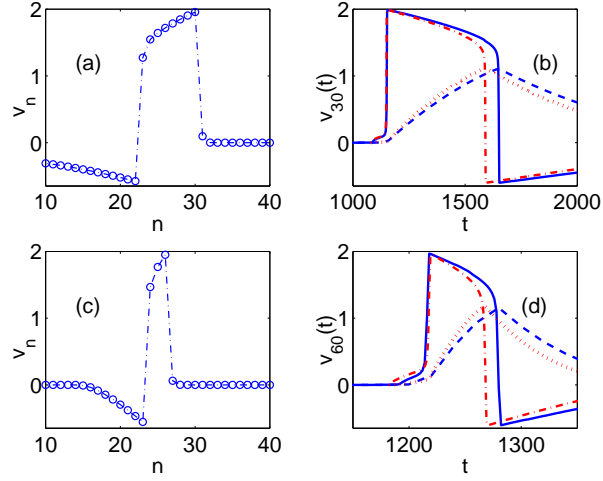}
\caption{Pulses  near $D_c$ and $\lambda_c(D)$:
(a) Spatial profile and (b) Temporal profile (solid line) when
 $D=0.004$ and $\lambda=0.0054$; (c) Spatial profile and
(d) Temporal profile (solid line) when $D=0.009$ and $\lambda=0.013$.
In (b) and (d), dashed lines represent the temporal profiles of
the slow variable. Dashed-dotted and dotted lines are the solutions
of the reduced equations (\ref{hd1})-(\ref{hd2}).
Other parameter values for FHN are $a=0.1$ and $b=0.5$.}
\label{figura5}
\end{center}
\end{figure}

\begin{figure}
\begin{center}
\includegraphics[width=8cm]{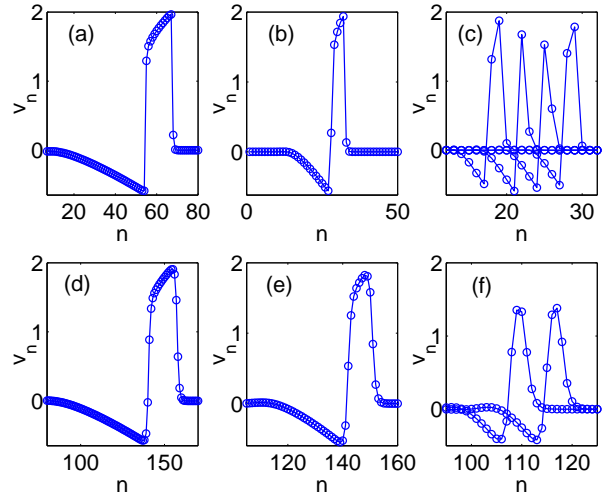}
\caption{Pulses for FHN with $a=0.1$,$b=0.5$ and increasing
$\lambda$. When $D=0.03$:
(a) $\lambda=0.01$, (b) $\lambda=0.03$,
(c) $\lambda=0.065$.
When $D=0.7$:
(d) $\lambda=0.05$, (e) $\lambda=0.1$,
(f) $\lambda=0.237$.}
\label{figura7}
\end{center}
\end{figure}

\begin{figure}
\begin{center}
\includegraphics[width=8cm]{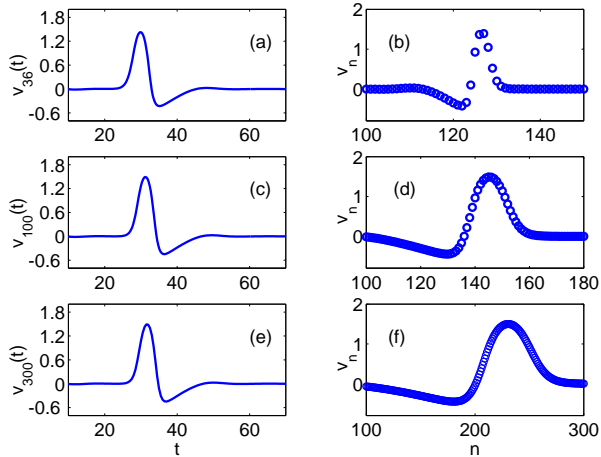}
\caption{Pulses for FHN near $\lambda_c(D)$ with $a=0.1$,$b=0.5$
and increasing $D$.
Temporal profiles when (a) $D=1$, (c) $D=10$, (e) $D=100$. The small
oscillation in the tail indicates that $(0,0)$ has just become a spiral
point.
Spatial profiles when (b) $D=1$, (d) $D=10$, (f) $D=100$.}
\label{figura11}
\end{center}
\end{figure}

In both discrete and continuous models, wave trains and pulses
disappear as $\lambda$ increases beyond certain thresholds.
However, the way in which these waves disappear can be rather
different.  In continuous models, the height of the peak of a
pulse or a wave train decreases as $\lambda$ increases and,
eventually, waves cease to propagate. In discrete models, new
modes of propagation failure arise depending on the value of
$D$. Unlike the continuous case, waves also fail to propagate
when $D$ is smaller than a critical value $D_c$. Figure \ref{figura8}
depicts the critical separation of time scales $\lambda_c(D)$
for FHN as a function of $D$.
We   distinguish three regions: small $D$
(highly discrete limit), large $D$ (continuum limit) and
intermediate $D$.
For the parameter values  selected in
Figures \ref{figura8}-\ref{figura11}, these regions are
approximately $D_c < D \leq 0.01$, $0.01 < D < 0.3$ and $0.3 \leq D $.
Figures \ref{figura8} (b) and (c) zoom in the highly discrete and
the intermediate regions.

Figure \ref{figura5} illustrates the shape of pulses for FHN when
$D$ is near $D_c$. For $\lambda$ close to $\lambda_c(D)$, the height
of the pulses has not decreased and remains near its maximum value
$2$. The width of the pulses does not vanish either. Moreover, it is
quite large near the critical coupling $D_c$. Close to
$\lambda_c(D)$, the peak contains $3$ points if $D=0.009$. This
number increases to $8$ points when $D=0.004$. For $D=0.003,$ we
reach $30$ points. The  number of points in the peak of a
propagating pulse near the critical time scale separation
$\lambda_c(D)$  seems to increase as $D$ approaches $D_c$. Then, why do
pulses fail to propagate? Near $D_c$ the leading edge advances in
the slow 'saltatory' manner typical of wave fronts at the
depinning transition \cite{cb}. Then, the separation of time scales
described in earlier Sections becomes blurred. Both variables, $v_n$
and $w_n$ change as the leading front moves. For $D$ fixed, and as
$\lambda$ increases, the leading edge becomes pinned and ceases to
propagate at a critical value of $\lambda$. Thus, in the highly
discrete region, waves fail to propagate at $\lambda_c(D)$  due to
the pinning of the leading edge.

For intermediate values of $D$, the peak of propagating pulses near the
critical time scale separation $\lambda_c(D)$ contains one point, see
Figure \ref{figura7} (c).
The temporal profiles of these pulses  show a peak with a narrow
trapezoidal  shape. Thus, their spatial profiles
change noticeably as the pulse moves, showing the configurations with
variable height depicted in Figure  \ref{figura7} (c). The reason
for propagation failure in this region is the vanishing width of the
peaks as $\lambda$ increases.

For larger $D$, the  mechanism of propagation failure as $\lambda$
grows is related to the decreasing height of the pulses,
as in the continuum limit. In fact, the profiles and speeds of the pulses
near $\lambda_c(D)$ resemble the profile and speed of the pulse solutions
of (\ref{cont}) near the critical time scale separation $\lambda_c$
for the continuous problem, as shown by Figures \ref{figura7} (f)
and \ref{figura11}.  As $D$ grows, more and more points accumulate
in the leading and trailing fronts (see Fig. \ref{figura11}) and
propagation fails before the top of the pulse is reduced to one point.

In this Section, we exploit (and correct) the predictions provided by the
asymptotic construction of travelling wave trains and pulses carried out in
Section \ref{sec:trains} to characterize the critical coupling $D_c$
and the critical separation of time scales for propagation $\lambda_c(D)$.
First, we consider travelling pulses and then we turn to wave trains.

\subsection{Propagation failure for pulses}

For a fixed $D$, the asymptotic construction in Section \ref{sec:pulses}
yields upper and lower bounds on the critical value $\lambda_c(D)$ above which
propagation of pulses fails, as a function of $D.$ Moreover, we find an asymptotic
prediction of  $\lambda_c(D)$ for small $D$, which agrees reasonably with numerical
measurements. Prior to this work, the only work in this direction is contained in the paper
by Chen and Hastings \cite{hastings}, who found rather coarse upper bounds on the
separation of time scales for a particular class of sources that excludes FHN and most
realistic sources.

An upper estimate  $\lambda_c(D)^+$ on
$\lambda_c(D)$ is found observing that our asymptotic
construction is consistent when the peak has one or more points:
$L_1(w^0,D,\lambda)\geq 1$. Therefore:
\begin{eqnarray}
\lambda \leq \lambda_c(D)  \leq  \lambda_c(D)^+ = c_+(w^0,D)
\int_{w^0}^{\overline{w}(w^0,D)}  {ds \over g(z_3(s),s)}. \label{p7}
\end{eqnarray}

On the other hand, numerical solutions show that pulses fail to propagate
when $L_1(w^0,D,\lambda)$ is smaller than a certain number of points
$k(D)$. Once $k(D)$ is determined:
\begin{eqnarray}
\lambda   \geq   \lambda_c(D)^-={c_+(w^0,D)  \over
k(D)} \int_{w^0}^{\overline{w}^0(w^0,D)}  {ds \over g(z_3(s),s)}
\label{p7bis}
\end{eqnarray}
yields a lower bound $\lambda_c(D)^-$ for $\lambda_c(D)$. Without
a theory to predict $k(D)$, our choice $k(D)=5$ for $D\in (0.01,1)$
in Figure \ref{figura8} (a) is empirical.

Figure \ref{figura8}(a) compares the numerically calculated
$\lambda_c(D)$ with the upper and lower bounds $\lambda_c(D)^{\pm}$
for the FitzHugh-Nagumo model.
The numerical approximation of $\lambda_c(D)$ is obtained as follows.
We solve the initial value problem (\ref{fh1})-(\ref{fh2}) starting
from the equilibrium state $(0,0)$ and excite the left boundary for a
large enough time $T$: $v_0(t)=2 \chi(0,T)$, $w_0(t)=0$, $t>0$.
If a pulse is successfully generated, we increase $\lambda$ slightly.
The process stops when we reach a value of $\lambda$ for which pulses
cease to be formed.

As Figure \ref{figura8}(a) shows, bounds (\ref{p7})-(\ref{p7bis})
are not sharp. There are several reasons for this.
First, the minimum number of points in the peak
before propagation failure changes with $D$. Second, as $\lambda$
grows the height of the propagating pulses diminishes and our
asymptotic construction loses precision. Third, near the critical
coupling $D_c$, the reduced bistable equation (\ref{p1}) does not
describe accurately the speed of the leading front. The relevance
of these three factors varies with $D$.  We will improve our predictions
of $\lambda_c(D)$ by a more precise study of the three regions
mentioned before: highly discrete, continuous and intermediate.

\subsubsection{Highly discrete limit}
\label{sec:hd}

In the highly discrete limit, $D\ll 1$, $w^0 \sim w_{c-}(D)$ and
we are near the depinning transition of wave fronts characterized by a
saltatory motion of active points, see Section \ref{sec:bistable}.
During the long time intervals between
abrupt jumps of the saltatory motion, $w_n$ can change appreciably
and the reduced equation for the motion of the leading edge of
pulses:
\begin{eqnarray}
{dv_n \over dt}= D (v_{n+1}-2v_n+v_{n-1}) +f(v_n,w^0)  \label{p1bis}
\end{eqnarray}
is no longer accurate. Instead of (\ref{p2bis}) with $w=w^0$,
the evolution of the active point is governed now by:
\begin{eqnarray}
{dv \over dt}= D(z_3(w)-2v+z_1(w)) +f(v,w),\label{hd1}\\
{dw \over dt}=\lambda g(v,w). \label{hd2}
\end{eqnarray}
Figures \ref{figura5} (b) and (d) compare  solutions of
(\ref{hd1})-(\ref{hd2}), represented by dashed lines, with the temporal
profile of the pulses. The leading edges are accurately described.

\begin{figure}
\begin{center}
\includegraphics[width=8cm]{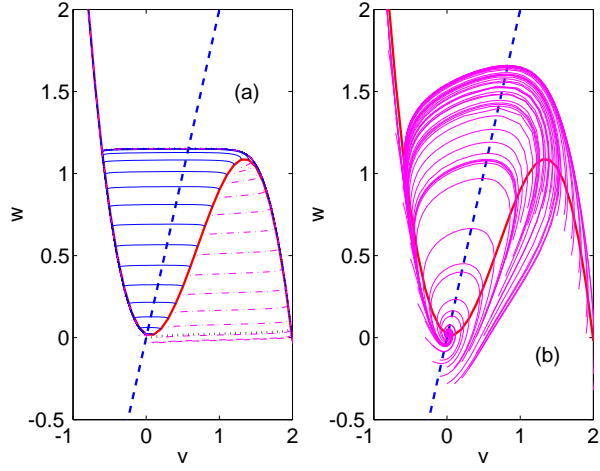}
\caption{Phase plane for the reduced  system (\ref{hd1})-(\ref{hd2}):
(a) $\lambda=0.01$, (b) $\lambda=0.1$. Parameter
values for FHN are $D=0.01$, $a=0.1$, $b=0.5$.}
\label{figura0}
\end{center}
\end{figure}

As we increase $\lambda$, we find a critical value $\lambda_0(D)$
at which trajectories of (\ref{hd1})-(\ref{hd2}) jumping from a
neighborhood of $(v^0,w^0)=(z_1(w^0),w^0)$ to a neighborhood of
$(z_3(w^0),w^0)$ cease to exist.
We obtain a first impression about $\lambda_0(D)$ by
examining the phase plane in Figure \ref{figura0}.
System (\ref{hd1})-(\ref{hd2}) has an equilibrium point.
For $\lambda$ small, the equilibrium point is a node and trajectories
starting at $(z_1(w^0),w^0)$ reach a neighborhood of
$(z_3(w^0),w^0)$, as shown by the  dashed lines in Figure
\ref{figura0} (a).  As $\lambda$ grows, it becomes a  spiral point.
Trajectories originating at $(z_1(w^0),w^0)$ wrap around the equilibrium
point, without reaching $(z_3(w^0),w^0)$,
see Figure \ref{figura0} (b). If we compute the value of
$\lambda$ at which the equilibrium changes type, we find an
approximate value for $\lambda_0(D)$. This approximation happens
to be quite poor: it is almost constant for $D$ small.

A better prediction of $\lambda_0(D)$ is found by calculating two
characteristic times. Suppose $w$ does not depart appreciably
from $w^0 \approx w_{c-}(D)$. The depinning analysis explained
in \cite{cb} shows that the quiescent period between successive jumps
in saltatory motion is $t_v \sim {\pi \over \sqrt{\alpha\beta
(w^0-w_{c-}(D))}}$. During this quiescent period, $v\sim s_0$,
$s_0$ being a value between $z_1(w^0)$ and $z_2(w^0)$ (see Section
2.1).
Suppose on the contrary that $\lambda$ is so large that $w$
increases from $w^0$ to $w_{c-}(D)$ while $v\sim s_0$. This occurs
in a time  $t_w=\int_{w^0}^{w_{c-}(D)} dw /\lambda g(s_0,w)$.
If $t_v \ll t_w$, saltatory motion of pulses can proceed.
In the opposite limit, $t_v \gg t_w$, $w$ evolves towards the
equilibrium value of (\ref{hd1})-(\ref{hd2}) and the pulse stops.
The crossover value at which $t_v\sim t_w$ yields the estimated
critical value $\lambda_0(D)$:
\begin{eqnarray}
\lambda_0(D) \sim {\sqrt{\alpha\beta (w^0-w_{c-}(D))} \over \pi}
\int_{w^0}^{w_{c-}(D)} {dw \over   g(s_0,w)}.
\label{hd3}
\end{eqnarray}
For FHN, $w^0=0$ and $\int_{0}^{w_{c-}(D)} {dw \over s_0-bw}
=-{1\over b} \ln(1-b{w_{c-}(D)\over s_0}) \sim {w_{c-}(D)\over s_0}$. Thus,
$\lambda_0(D) \sim {w_{c-}(D)\over s_0}
{\sqrt{\alpha\beta (-w_{c-}(D))}\over \pi}$.
Figure \ref{figura8} (b) compares $\lambda_c(D)$ with the numerically
measured value of $\lambda_0(D)$, our asymptotic prediction (\ref{hd3})
and the upper bound $\lambda^+(D)$. Now, this upper bound is computed using
the approximation of the speed provided by  (\ref{hd1})-(\ref{hd2})
instead of the speed of the wave front solution of (\ref{p1}).
Notice the good agreement between $\lambda_c(D)$ and $\lambda_0(D)$
for small $D$.

\subsubsection{Intermediate region and Continuous limit}

In the intermediate region,  the peak of the pulses contains one
point near $\lambda_c(D)$ and the upper bound $\lambda^+(D)$
provides a good estimate of the critical threshold.
Formula (\ref{p7}) is more accurate
when we replace the speed of the wave front
solutions of the reduced bistable equation (used in
Figure \ref{figura8} (a)) by the speed found
solving numerically (\ref{hd1})-(\ref{hd2}) (used in
Figure \ref{figura8} (c)).

For larger $D$, the upper and lower bounds $\lambda^{\pm}(D)$ yield
poor approximations due to the decreasing height of pulses.
As Figure \ref{figura8} (a) shows,  $\lambda_c(D)$ increases slowly
to the threshold $\lambda_c$ for the continuous problem.
Notice that equations (\ref{int1})-(\ref{int2})
with $D(h)={1\over h^2}$ are the method of lines for the numerical
approximation of the solutions of the continuous system
(\ref{cont}) as $h\rightarrow 0$.
For large $D$, the continuum limit yields good approximations
to thresholds and pulse profiles.
It is well known that (\ref{cont}) has two families of pulses:
A fast (stable) family and a slow (unstable) family.
For discrete models with piecewise
linear sources, fast and slow families of pulses were constructed
by Tonnelier \cite{tonnelier}. At a critical
$\lambda$ both families collapse and disappear.
This suggests that a similar bifurcation
 might take place at $\lambda_c(D)$ for general
 discrete models.

\subsection{Propagation failure for wave trains}

The asymptotic construction of wave trains in Section \ref{sec:asymptotics}
is restricted to the range $w \in (w^0,w_{c-}(D))$. For $D$ fixed, wave
trains die out as $w \rightarrow w_{c-}(D)^-$ due to the pinning  of the
leading wave front. The speed of the wave train  can be approximated by
the formula $c\sim \sqrt{\alpha(D)\beta(D) (w-w_{c-}(D))}/\pi$, obtained in
\cite{cb} by the depinning analysis of wave fronts in discrete bistable
equations. When $w$ is near the thresholds $w_{c\pm}(D)$, the profiles
develop steps as shown by Figures \ref{figura1} and \ref{figura5}. Propagation
is 'saltatory'.
As $D$ decreases, the range $(w^0,w_{c-}(D))$ for which wave trains
exist becomes smaller, and it disappears at the critical coupling
$D_c$ such that $w_{c-}(D_c)=w^0$.

Our asymptotic construction also yields  estimates on the critical values
of $\lambda$ above which propagation of trains fails.
For each $w\in (w^0,w_{c-}(D))$,
there is a threshold $\lambda_c(w,D)$ below which wave trains with
speed $c_+(w,D)$ are found. An upper estimate  $\lambda_c(w,D)^+$ on
$\lambda_c(w,D)$ is obtained observing that our asymptotic
construction is consistent when the peak has one or more points:
$L_1(w,D,\lambda)\geq 1$. Therefore:
\begin{eqnarray}
\lambda \leq \lambda_c(w,D)  \leq  \lambda_c(w,D)^+ = c_+(w,D)
\int_{w}^{\overline{w}(w,D)}  {ds \over g(z_3(s),s)}. \label{p8}
\end{eqnarray}
The largest value of $\lambda_c(w,D)^+$ is attained for pulses:
$\lambda_c(D)^+=\lambda_c(w^0,D)^+$. A lower bound is found proceeding in a
similar way as we did for pulses.

\section{Relaxation oscillations}
\label{sec:oscillatory}

\begin{figure}
\begin{center}
\includegraphics[width=8cm]{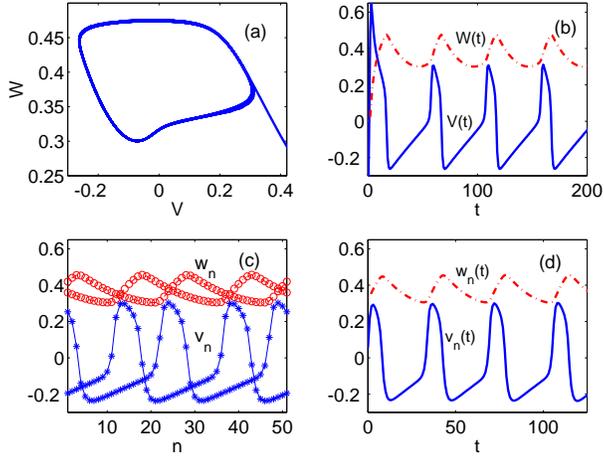}
\caption{For ML:
(a) Limit cycle and (b) Periodic trajectory for an uncoupled unit;
(c) spatial profiles at different times: $v_n(t)$ (asterisks), $w_n(t)$
(circles); (d) temporal profiles of $v_n(t)$ (solid line) and $w_n(t)$
(dashed line) for $n=20$ fixed. Parameter values are $I=0.375$, $D=0.5$,
$\lambda=0.01$.}
\label{figura9}
\end{center}
\end{figure}

When the nullclines of (\ref{int1})-(\ref{int2}) intersect at an
unstable equilibrium $(v^0,w^0)$ located in the central branch of the cubic,
excitability is lost and spatially uniform  profiles correspond to
a stable limit cycle of  system  (\ref{ode}) for $\lambda\leq \lambda_c$.
Let $V(t)$ and $W(t)$ denote this limit cycle. Its period $T>0$,
increases proportionally to $\lambda^{-1}$ as $\lambda \rightarrow 0$.
Figures \ref{figura9}(a)-(b) illustrate  the structure of the limit cycle.

In this case, periodic wave train solutions of (\ref{int1})-(\ref{int2})
are generated using scaled periodic configurations $v_n(0)=V(n/c)$,
$w_n(0)=W(n/c)$ as initial data, for any $c>0$.
For continuous systems, this fact was pointed out by Maginu
\cite{maginu2}.
Figures \ref{figura9} (c)-(d) show a periodic wave train for  $c=2$.
We find wave trains with spatial period $T/c$ and speed $c$. In contrast
with the excitable case, such wave trains cannot be generated from an
equilibrium state by periodic excitation of a boundary.

\begin{figure}
\begin{center}
\includegraphics[width=8cm]{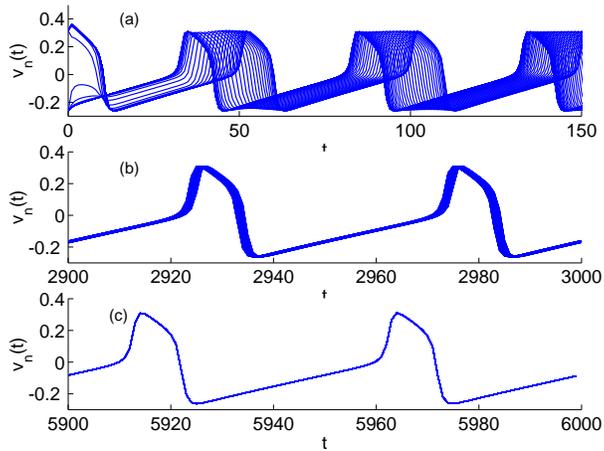}
\caption{Synchronization in the  Morris-Lecar model. We have
superimposed the time evolution of $N=50$ oscillators. Parameter
values are $I=0.375$, $D=0.5$ and $\lambda=0.01$.}
\label{figura10}
\end{center}
\end{figure}

More often, we expect nonuniform profiles $v_n(t)$ and $w_n(t)$ to
display synchronization phenomena.
Figure \ref{figura10} illustrates synchronization of a population of
$N=50$ oscillators with zero Neumann boundary conditions. Initially,
half the population is synchronized at a certain phase and the other half
at a different phase.
The trajectories $v_n(t)$, $w_n(t)$ of individual points  behave
as  $V(t+\phi_n),W(t+\phi_n)$, with a slowly varying phase $\phi_n$ which
may become independent of $n$ as $t\rightarrow \infty$, as shown in Figure
\ref{figura10}. Notice that the diffusive coupling is only  active in the
part of the cycle where $v_n$ changes abruptly.

For a better understanding of the evolution of different initial configurations
we need an equation for the phases $\phi_n(t)$. We will show below that,
for small couplings $D \ll 1$, the oscillator phases $\phi_n(t)$ obey the
equation:
\begin{eqnarray}
{d \phi_n \over d\tau}= {1\over T}\!\!\int_0^T \!\!\! y_1(\theta_n)
[V(\theta_n+\phi_{n+1}-\phi_n)-2V(\theta_n)+V(\theta_n+\phi_{n-1}-\phi_n)]
d \theta_n
\label{syn1}\end{eqnarray}
where $\theta_n=t+\phi_n(\tau)$ and $(y_1(\theta),y_2(\theta))$ is a
T-periodic solution of the adjoint homogeneous problem:
\begin{eqnarray}
\left(  \begin{array}{c} y_{1,t}  \\ y_{2,t} \end{array} \right)   +
\left(  \begin{array}{cc} f_v(V,W) & \lambda
g_v(V,W) \\ f_w(V,W) & \lambda g_w(V,W) \end{array} \right)
\left(  \begin{array}{c} y_1  \\ y_2 \end{array} \right) =
\left(  \begin{array}{c} 0  \\ 0 \end{array} \right). \label{syn2}
\end{eqnarray}
normalized by ${1\over T} \int_0^T
y_1(\theta)V'(\theta)+y_2(\theta)W'(\theta)=1$. For small $\lambda$,
system (\ref{syn2}) decouples and $y_1$ is computed from $y_{1,t} +
f_v(V,W) y_1=0$ with  ${1\over T} \int_0^T y_1(\theta)V'(\theta)=1.$
In general, $V$ and $W$ are not known explicitly. However, an
asymptotic approximation is found using the separation of the time
scales. The profiles are composed of regions where $V$ varies
rapidly in the time scale $t$ whereas $W$ is almost constant and
regions where $W$ varies slowly in the time scale $\lambda t$
whereas $V$ is at equilibrium, $f(V,W)=0$.

Equation (\ref{syn1}) is obtained as follows. Set $\tau=Dt$.
Inserting the expansions $v_n=V(t+\phi_n(\tau))+ D v^{(1)}_n(t,\tau) +O(D^2),$
$w_n=W(t+\phi_n(\tau))+ D w^{(1)}_n(t,\tau) +O(D^2)$ in (\ref{int1})-(\ref{int2})
we get:
\begin{eqnarray}\begin{array}{r}
 D v^{(1)}_{n,t}  + (1+ D \phi_{n,\tau})V' =
f(V,W) + D [f_v(V,W) v^{(1)}_n + f_w(V,W) w^{(1)}_n  ] \\
+ D [V(t+\phi_{n+1}(\tau))-2V(t+\phi_n(\tau))+V(t+\phi_{n-1}(\tau))]+ O(D^2)
\end{array} \label{syn3}\\
\begin{array}{r}
 D w^{(1)}_{n,t}  +  (1+ D \phi_{n,\tau})W'  =
\lambda g(V,W) + \lambda D [g_v(V,W) v^{(1)}_n + g_w(V,W) w^{(1)}_n] \\+O(D^2).
\end{array} \label{syn4}
\end{eqnarray}
This yields:
\begin{eqnarray}\begin{array}{l}
 v^{(1)}_{n,t}- [f_v(V,W) v^{(1)}_n + f_w(V,W) w^{(1)}_n ] =-
\phi_{n,\tau}V'(t+\phi_n(\tau))
\\+ [V(t+\phi_{n+1}(\tau))-2V(t+\phi_n(\tau))+V(t+\phi_{n-1}(\tau))]\\
w^{(1)}_{n,t}- \lambda [g_v(V,W) v^{(1)}_n + g_w(V,W) w^{(1)}_n]=- \phi_{n,\tau}
W'(t+\phi_n(\tau)) \label{syn5}
\end{array}\end{eqnarray}
The solvability condition for system (\ref{syn5}) is precisely our equation
(\ref{syn1}). Equations for the time evolution of the oscillator phases
in continuous problems were derived in \cite{neu2,kuramoto}.
For discrete problems with diffusion in both the fast and slow variables see
\cite{neu}.


\section{Conclusions}

\label{sec:conclusions}

We have studied wave propagation and oscillatory behavior in
excitable and self-oscillatory one dimensional systems with one fast
and one slow variables. In excitable systems, we have constructed
asymptotically a one-parameter family of stable  wave trains with
increasing periods and speeds. Our numerical experiments suggest
that the stability of wave trains in discrete systems can be
characterized analytically in terms of the dispersion law, as it
happens for wave trains in continuous systems \cite{maginu}. When
the spatial period tends to infinity, we obtain the fastest wave of
the family: a solitary pulse. As the strength of the coupling
between nodes decreases, the family of wave trains becomes smaller.
The disappearance of the solitary pulse marks the onset of
propagation failure due to spatial discreteness. The critical
coupling is the pinning threshold for the leading edge of the pulse
in a reduced bistable equation. Close to failure, the wave profiles
develop steps and propagation becomes saltatory. Instead of
propagating smoothly, the excitation jumps from node to node.
Another source of propagation failure is the small separation of
time scales for excitation and recovery, shared by continuous
models. We obtain bounds for the critical separation of time scales
as a function of the coupling. Technically, we have greatly improved
the rough bounds contained in \cite{hastings}, derived under very
restrictive assumptions on the nonlinearities, which exclude FHN and
most realistic sources. Such bounds are unknown in the continuum
case. For small $D$, we also develop an asymptotic theory of failure
as $\lambda$ decreases. Our asymptotic predictions show a good
agreement with numerical solutions of FHN and ML models.

In self-oscillatory systems, wave trains are generated choosing
scaled periodic configurations as initial data. For other initial
configurations, we observe synchronization phenomena. In the limit
of small coupling, we find an equation for the time evolution of the
oscillator phases. A detailed analysis of this equation which would
support the numerical evidence of synchronization remains to be
done.

\vskip 5mm {\bf Acknowledgments.} This work has been supported by
the Spanish MCyT through grant BFM2002-04127-C02, and by the
European Union under grant HPRN-CT-2002-00282.
The author thanks Prof. L.L. Bonilla for fruitful discussions.

\end{document}